\documentclass[aps,prl,nofootinbib,superscriptaddress,twocolumn]{revtex4}
\usepackage{latexsym}
\usepackage{amsthm}
\usepackage{amsmath}
\usepackage{graphicx}
\usepackage{amssymb}
\usepackage{hyperref}
\usepackage{bbm}
\newcommand{\gsim}{\lower.7ex\hbox{$\;\stackrel{\textstyle>}{\sim}\;$}}
\newcommand{\lsim}{\lower.7ex\hbox{$\;\stackrel{\textstyle<}{\sim}\;$}}
\def\OO{{\cal O}}
\def\LL{{\cal L}}
\def\PP{{\cal P}}

\newcommand{\rel}{{\text{rel}}}

\newcommand{\GeV}{\,\mathrm{GeV}}

\newcommand{\keV}{\,\mathrm{keV}}

\newcommand{\half}{{\frac{1}{2}  }}

\newcommand{\slD}{D\!\!\!\!\!\!\!\not\;\;\;}

\newcommand{\Tr}{{\text{ Tr }}}

\newcommand{\eff}{{\text{eff}}}

\newcommand{\be}{\begin{equation}}
\newcommand{\ee}{\end{equation}}
\newcommand{\bea}{\begin{eqnarray}}
\newcommand{\eea}{\end{eqnarray}}

\newcommand{\bef}{\begin{figure}[htbp]\begin{center}}
\newcommand{\eef}{\end{center}\end{figure}}
\newcommand{\ket}[1]{|#1\rangle}
\newcommand{\bra}[1]{\langle #1|}

\newcommand{\pid}{{\pi_{\text{d}}}}
\newcommand{\rhod}{{\rho_{\text{d}}}}

\begin{document}

\pagestyle{plain}

\title{
\begin{flushright}
\mbox{\normalsize SU-ITP-09/13}
\end{flushright}
\vskip 15 pt

Composite Inelastic Dark Matter }

\author{Daniele S. M. Alves}
\affiliation{
SLAC, Stanford University, Menlo Park, CA 94025}
\affiliation{
Physics Department, Stanford University,
Stanford, CA 94305}
\author{Siavosh R. Behbahani}
\affiliation{
SLAC, Stanford University, Menlo Park, CA 94025}
\affiliation{
Physics Department, Stanford University,
Stanford, CA 94305}
\author{Philip Schuster}
\affiliation{
SLAC, Stanford University, Menlo Park, CA 94025}
\author{Jay G. Wacker}
\affiliation{
SLAC, Stanford University, Menlo Park, CA 94025}
\date{\today}
\begin{abstract}
Peaking consistently in June for nearly eleven years, 
the annual modulation signal reported by DAMA/NaI and DAMA/LIBRA offers strong evidence for the identity of dark matter.
DAMA's signal strongly suggest that dark matter inelastically scatters into an excited state split by $\OO(100 \keV)$. 
We propose that DAMA is observing hyperfine transitions of a composite dark matter particle. 
As an example, we consider a meson of a QCD-like sector, built out of constituent fermions whose spin-spin interactions break the degeneracy of the ground state.  
An axially coupled $U(1)$ gauge boson that mixes kinetically with hypercharge
induces inelastic hyperfine transitions of the meson dark matter that can explain the DAMA signal.
\end{abstract}

\maketitle


This letter proposes a new class of inelastic dark matter (iDM) models  that can explain
the annual modulation reported by DAMA/NaI and DAMA/LIBRA \cite{Bernabei:2005hj}. 
DAMA's signal peaks in early June, consistent with dark matter scattering, and has remained in phase for nearly eleven years. 
Moreover, the fractional modulation of the signal appears anomalously large, and the nuclear recoil spectrum has a peak near $E_R \simeq \OO(30 \keV)$. 

The hypothesis that dark matter scatters inelastically off nuclei into a $\OO(100 \keV)$ excited state elegantly explains the salient features of the DAMA signal \cite{TuckerSmith:2001hy}.  IDM models predict nuclear recoil spectra with a characteristic peak and an $\OO(1)$ modulation fraction \cite{Weiner:2008}. 
The large dark matter velocity threshold required by inelastic kinematics also implies that heavier nuclei targets like ${}^{127}\text{I}$ in DAMA provide enhanced signal sensitivity relative to lighter targets such as ${}^{74}\text{Ge}$ in CDMS.

In composite inelastic dark matter models (CiDM),  DAMA's observed signal arises from inelastic hyperfine transitions of a composite dark matter particle.  
(For other examples of composite dark matter, see \cite{Technibaryon}.)
We illustrate this mechanism with a simple model where the majority of dark matter is a meson of a strongly coupled $SU(N_c)$ gauge  sector that confines near $\Lambda\simeq \GeV$. 
These mesons are comprised of constituent fermions whose hyperfine interactions split the ground state by $\OO(100 \keV)$.  
When one constituent quark is non-relativistic, a hierarchy between the hyperfine scale and the dark matter mass follows inevitably from an enhanced spin symmetry.
The dark matter couples to a new $U(1)_{A'}$ vector boson that kinetically mixes with the Standard Model's hypercharge\cite{Holdom:1985}.  Another version of iDM with kinetic mixing is given in \cite{ArkaniHamed:2008}.
Axial couplings of the $U(1)_{A'}$ to the constituent fermions mediate inelastic hyperfine transitions 
that dominate low-energy nuclear scattering. 

The model considered here has two Dirac fermions, $\Psi_H$ and $\Psi_L$, transforming in the fundamental representation of the $SU(N_c)$ gauge group. 
The new $U(1)_{A'}$ couples axially to $\Psi_{H,L}$, each of which have equal and opposite unit charge. 
We introduce a charge-2 Higgs $\phi$, whose vacuum expectation value generates a mass for $\Psi_{H,L}$ and the $A'$.
The $U(1)_{A'}$ is non-anomalous for this particle content.
The dark matter candidate is a $\bar{\Psi}_L \Psi_H$ bound state, and its stability can be guaranteed by imposing a $U(1)_{H-L}$ flavor symmetry or $\mathbb{Z}_2$ under which $\Psi_H\rightarrow -\Psi_H$. The Lagrangian for this model is
\begin{eqnarray}
\LL &=& \LL_{\text{SM}} + \LL_{\Psi}+\LL_{\text{Gauge}}+ \LL_{\text{break}};\\
\LL_{\text{Dark Gauge}}&=&- \frac{1}{2}\!\! \Tr G_{\mu\nu}^2\!  -\! \frac{1}{4} F_{A'}^2 + \epsilon F_{A'}^{\mu\nu}B_{\mu\nu} \label{eq:mix}.
\\ 
\nonumber
\LL_{\Psi} &=& \bar{\Psi}_L i \slD \Psi_L + 
y_L  \bar{\Psi}_L (\phi \PP_L + \phi^* \PP_R) \Psi_L
 \\
 \nonumber
&&\hspace{-0.05in}+ \bar{\Psi}_Hi \slD \Psi_H + y_H  \bar{\Psi}_H( \phi^*\PP_L  + \phi\PP_R )\Psi_H\\
\nonumber
 \LL_{\text{break}} &=& |D_\mu \phi|^2 - \lambda( |\phi|^2 - v_\phi^2)^2,
\end{eqnarray}
where $\PP_{L,R}$ are the left and right Dirac projection operators, and
 $D_\mu\Psi_{L,H} = (\partial_{\mu}+ i g_D G_\mu\pm ie' \gamma_5 A'_{\mu} )\Psi_{L,H}$ where $G_\mu$ is the $SU(N_c)$ gauge field and $\pm$ corresponding to $\Psi_H$ and $\Psi_L$, respectively.  
 $B^{\mu\nu}$ is the hypercharge field strength, $F^{\mu\nu}_{A'}$ is the $U(1)_{A'}$ field strength, and $G_{\mu\nu}$ is the field strength for the confining $SU(N_c)$ theory.
In this model, the dark quarks have a hierarchy of masses $m_L= y_L v_\phi \leq \Lambda \ll m_H= y_H v_\phi$, and an $A'$ mass in the range $100 \text{ MeV}\lsim m_{A'} \lesssim 20 \text{ GeV}$.  

At momenta beneath $\Lambda$, the model consists of mesons and baryons built out of $\Psi_H$ and $\Psi_L$, listed below.
Because $m_H\gg m_L$, we  classify the states by the number of  $\Psi_H$ constituents, $N_H$.  
$N_H$ $\Psi_H$ can form bound states by anti-symmetrizing their color indices.   They have binding energies
\begin{eqnarray}
E_B \propto  \alpha_t^2(\mu^*)m_H/ N_c^2 \label{eq:binding}
\end{eqnarray}
where $\alpha_t(\mu^*)$ is the running 't Hooft coupling of the strong gauge sector evaluated at the inverse Bohr radius 
of the bound state, $\mu^*\simeq\alpha_t m_H$.
At distances greater than $\Lambda^{-1}$, the color charge of $N_H$ heavy $\Psi_H$'s can be screened by
$N_H$ light $\bar{\Psi}_L$ antiquarks to form a dark $N_H$-meson, or by $N_c-N_H$ $\Psi_L$ quarks to form a dark $N_H$-baryon.
The $\Psi_H$ and $\Psi_L$ quarks have antisymmetrized colors, so the spins of same-flavor constituents must be symmetrized.
The resulting range of spins for dark mesons and baryons is
\begin{eqnarray}
0\le j_{N_H \text{M}}\le N_H\qquad 
\Big|\half N_c -N_H\Big|\le  j_{N_H \text{B}}\le \half N_c  .
\end{eqnarray}
Due to the spin-spin interactions, the lowest-spin configurations with a given $N_H$ are the lowest-energy configurations.   
In particular, all dark mesons have a spin-0 ground state. 

Cosmology dominantly produces $N_H=1$, $J^P=0^-$ mesons $\pid=\bar{\Psi}_L\gamma_5\Psi_H$, as will be shown later. 
In the limit $m_L\leq \Lambda\ll m_H$,  
the $J^P=1^-$ vector $\rhod =\bar{\Psi}_L \gamma_\mu \Psi_H$ is nearly degenerate with $\pid$, 
and is accessible in low-energy $\pid$ scattering.
Spin-spin interactions generate a mass splitting
\begin{eqnarray}
\label{Eq: Delta}
\Delta\equiv M_{\rhod}-M_{\pid}=\frac{\kappa\Lambda^2}{M_\pid},
\end{eqnarray}
where $M_{\pid}\simeq m_H$ is the $\pid$ mass, and $\kappa$ is an order-unity 
coefficient that  fixes the relation between $\Delta$ and $\Lambda$. 
The $\pid$ and $\rhod$ mesons form a multiplet of the $SU(2)_{H\text{-spin}}$ that rotates $\Psi_H$\!'s spin. 
In the limit $m_H\gg \Lambda$, $SU(2)_{H\text{-spin}}$ is an approximate symmetry, guaranteed by Lorentz invariance.
This is a familiar phenomenon in heavy-quark physics and the enhanced symmetry constrains low energy 
interactions of $\pid$ and $\rhod$ mesons \cite{Georgi:1990um}.

\section{CiDM Scattering}
The low-energy scattering of the dark mesons arises after diagonalizing
the kinetic mixing terms in (\ref{eq:mix}), and integrating out the weak interactions.
The dark matter constituents couple to the Standard Model via
\begin{eqnarray}
\mathcal L_{\text{Int}}\! =\! \frac{-\epsilon s_ \theta }{m_{Z^0}^2} J_{Z^0\!\mu} J_{\text{d}}^{\mu} 
+\! \! \left(\!J_{\text{d}}^{\mu}\!- \epsilon c_{\theta} J_{\text{EM}}^{\mu}\! - \epsilon s_{\theta} \frac{ m_{A'}^2  }{m_{Z^0}^2}J_{Z^0}^{\mu}\!\!\right)\!\!
{A'}_{\mu}   ,\label{eq:SMinteractions}
\end{eqnarray}
where $s_\theta= \sin\theta_{\text{w}}$ and  $c_\theta= \cos\theta_{\text{w}}$, $J_{\text{EM}}$ is the electromagnetic current, $J_{Z^0}$ is the neutral $Z^0$  current and $J_{\text{d}}$ is the dark sector $U(1)_{A'}$ current,
\begin{eqnarray}
J_{\text{d}}^\mu =  \bar{\Psi}_H \gamma^\mu\gamma_5\Psi_H - \bar{\Psi}_L \gamma^\mu\gamma_5\Psi_L. \label{eq:current} 
\end{eqnarray}
As discussed in \cite{Lisanti:2009am}, parity-breaking by strong dynamics in the dark sector 
induces elastic charge-radius $\pid +\text{SM}\rightarrow \pid +\text{SM}$ scattering, which has new  
consequences for direct detection.
For simplicity, we will discuss the case where the strong dynamics preserves parity in the dark sector, 
so that no $\pid +\text{SM}\rightarrow \pid +\text{SM}$ elastic scattering is mediated by the $U(1)_{A'}$ current 
in (\ref{eq:current}) to $\OO(\epsilon)$.
So long as $\phi$ does not significantly mix  with the Standard Model Higgs boson, 
elastic dark-matter-nuclues scattering induced by $\phi$ exchange is also negligible.

Using heavy quark effective theory \cite{Georgi:1990um}, the $SU(2)_{H\text{-spin}}$ and 
Lorentz symmetry of the $\pid, \rhod$ mesons constrain the form of their scattering matrix elements. 
At leading order in relative velocity, $v_\rel/c\simeq 10^{-3}$, the $\pid\rightarrow \rhod$ matrix elements are given by,
\bea
\bra{\rhod(p'\!,\epsilon)}J^{\mu}_{\text{d}}\ket{\pid(p)}\!\! \simeq\! 4M_\pid\epsilon^{\mu}_{p'}
\!+\! \frac{c_{\text{in}}}{\Lambda}(p+p')^\mu\! q_\nu\epsilon^\nu_{p'}, \label{eq:SpinDipole}
\eea
where $\epsilon^{\mu}_{p'}$ is the polarization of $\rhod$, and $c_{\text{in}}\sim 1$ controls dipole scattering. 
We have dropped terms proportional to $q^{\mu}$, which have vanishing contraction with the conserved 
Standard Model electromagnetic current.
The second term in (\ref{eq:SpinDipole}) leads to scattering enhanced by $q^2/v_\rel^2\Lambda^2\simeq 10^3$ relative to the first term.

In terms of relativistic effective operators, the low energy interactions can be described as:
\bea
\LL_\text{eff}&=&d_\text{in}M_\pid \pid^{\dagger}\rhod^\mu A'_\mu+\frac{c_\text{in}}{\Lambda}\pid^{\dagger}\partial_\mu\rhod_\nu F_{A'}^{\mu\nu} \\
\nonumber
&&+\frac{d_\text{el}}{\Lambda^2}\partial_\mu\pid^\dagger\partial_\nu\pid\tilde{F}_{A'}^{\mu\nu}+...
\eea

Elastic transitions mediated by the $d_\text{el}$-operator above are velocity suppressed relative to the inelastic ones. The  axial coupling of $A'$ to the fermionic constituents leads to a parity constraint on the interactions, forbidding elastic operators of the type
\be
\mathcal{O}_\text{forbidden}=\frac{c_\text{el}}{\Lambda^2}\partial_\mu\pid^\dagger\partial_\nu\pid F_{A'}^{\mu\nu}
\ee
that if otherwise allowed would dominate over inelastic scattering.

%

Using (\ref{eq:SpinDipole}), the low-energy inelastic cross-section for $\pid$ to scatter off a nucleus of mass $m_N$ and electric charge $Z$ with nuclear recoil energy $E_R$ is
\bea
\frac{d\sigma(E_R,v_\rel)}{dE_R} \simeq \frac{ 4 Z^2\alpha  }{M_\pid \Delta \;f_\eff^4}\;\frac{m_N^2E_R  |F_{\text{H}}(E_R)|^2}{v_\rel^2(1 +2m_N E_R/m_{A'}^2)^2}, \label{eq:SpinDipoleScattering}
\eea
where
\begin{eqnarray}
\label{Eq: feff}
f^4_\eff = m_{A'}^4/ (c_{\text{in}}\epsilon c_\theta g_{\text{d}})^2 \kappa  
\end{eqnarray}
and $F_{\text{H}}$ is the Helm nuclear form factor used in \cite{Lewin:1995rx}. 
The differential scattering rate per unit detector mass is
\begin{eqnarray}
\frac{dR}{dE_R} =  
\frac{\rho_0 v_0}{m_N M_\pid}
\int_{v_{\text{min}}(E_R)}
\hspace{-0.3in}d^3 v_\rel \hspace{0.1in} 
\frac{v_\rel}{v_0} f(v;v_e) \frac{d\sigma}{dE_R},
\end{eqnarray}
where $\rho_0=0.3 \GeV/\text{cm}^3$ is the local density of dark matter, $v_{\text{min}}(E_R)$ is the minimum relative velocity required to scatter with nuclear recoil energy $E_R$, and $f(v;v_e)$ is the dark matter velocity distribution function in the lab frame accounting for the Earth's variable velocity $\vec{v}_e$.
Naively cutting off the velocity profile above the galactic escape velocity $v_{\text{esc}}$, we use the Standard Halo Model velocity distribution function,
\bea
f(v;v_e) \propto \Big(e^{ - \frac{(\vec{v} - \vec{v}_e)^2}{ v_0^2}}\! - e^{- \frac{v_{\text{esc}}^2}{v_0^2}} \Big)
 \Theta(v_{\text{esc}}-|\vec{v}-\vec{v}_e|).
\eea

Given the high uncertainty on the halo velocity distribution and the sensitivity of iDM models to it \cite{Alves:2010pt}, we marginalize over the velocity parameters $v_0$ and $v_{\text{esc}}$, constrained to be in the range $150~\text{km/s}\leq v_0 \leq 350~\text{km/s}$ and $480~\text{km/s}\leq v_{\text{esc}}\leq650~\text{km/s}$.

\section{Direct Detection}

\begin{table*}[htdp]
\label{NullData}
\begin{center}
\begin{tabular}{|c||c|c|c|c|c|c|}
\hline
Experiment & Element & Reference & Effective exposure & Period of run & Signal Window & Obs. events\\ 
\hline \hline
  CDMS '05 & Ge & \cite{Akerib:2005kh} & 34 kg-d & 2005 Mar 25 - 2005 Aug 8 & 10 -100 keV & 1\\ 
 \hline
 CDMS '07 & Ge & \cite{Ahmed:2008eu} & 121.3 kg-d & 2006 Oct 1 - 2007 Jul 1 & 10 -100 keV & 0\\ 
 \hline
 CDMS '08 & Ge & \cite{Ahmed:2009zw} & 194.1 kg-d & 2007 Jul 1 - 2008 Sep 1 & 10 -100 keV & 2\\ 
 \hline
 XENON10 & Xe & \cite{Angle:2007uj} & 0.3 $\times$ 316.4 kg-d & 2006 Oct 6 - 2007 Feb 14 & 4.5 -75 keV & 13\\ 
 \hline
 CRESST-II '04 & W & \cite{Angloher:2004tr} & 0.59 $\times$ 0.9 $\times$ 20.5 kg-d & 2004 Jan 31 - 2004 Mar 23 & 12 -100 keV & 5\\ 
 \hline
CRESST-II '07 & W & \cite{Angloher:2008jj} & 0.59 $\times$ 0.9 $\times$ 48 kg-d & 2007 Mar 27 - 2007 Jul 23 & 12 -100 keV & 7\\ 
 \hline
ZEPLIN-II & Xe & \cite{Alner:2007ja} & 0.5 $\times$ 225 kg-d & 2006 May 1 - 2006 Jun 30 & 13.9 -55.6 keV & 29\\ 
 \hline
ZEPLIN-III & Xe & \cite{Lebedenko:2008gb} & 0.5 $\times$ 126.7 kg-d & 2008 Feb 27 - 2008 May 20 & 10.7 -30.2 keV & 7\\ 
 \hline
ZEPLIN-III (iDM) & Xe & \cite{Akimov:2010vk} & 0.5 $\times$ 63.3 kg-d & 2008 Feb 27 - 2008 May 20 & 17.5 -78.8 keV & 5\\ 
 \hline
 XENON100 & Xe & \cite{Aprile:2010um} & 161 kg-d & 2009 Oct 20 - 2009 Nov 12 & 7.4 -29.1 keV & 0\\ 
 \hline
 
\end{tabular}
\caption{Summary of the data from null experiments.}
\end{center}
\end{table*}%

In DAMA, dark matter dominantly scatters off ${}^{127}\text{I}$. The modulation spectrum and rate of the combined 1.17 ton-yr exposure \cite{Bernabei:2005hj} constrain the parameters in this model. We perform a global $\chi^2$ fit by marginalizing over the 5 unknown parameters: $M_\pid$, $\Delta$, $f_{\text{eff}}$, $v_0$ and $v_{\text{esc}}$. 

\begin{figure}[htbp]
\includegraphics[width=3.4in]{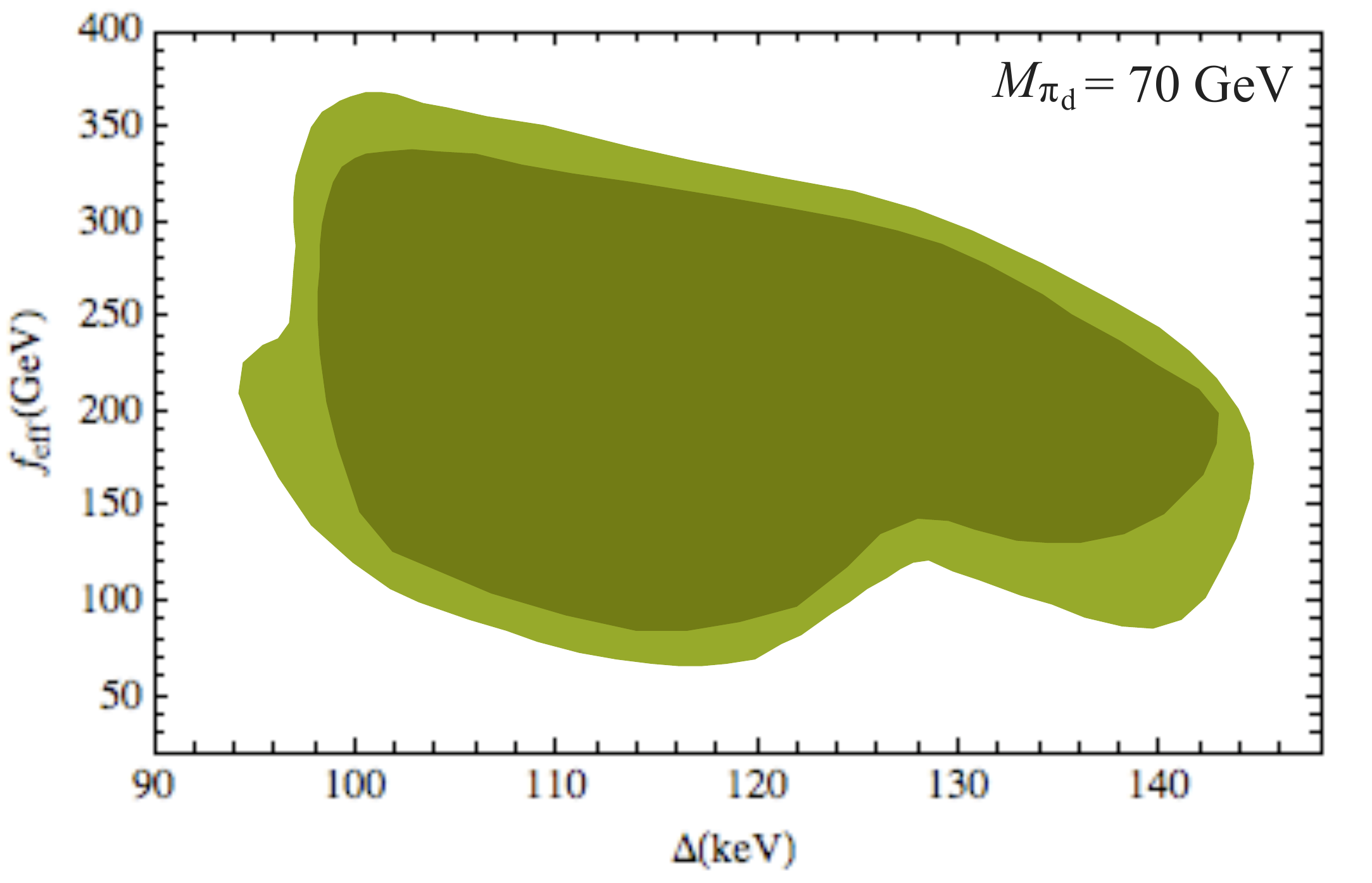}
\caption{\label{Fig: Banana} Values of $f_{\eff}$, defined in (\ref{Eq: feff}),  and $\Delta$, defined in (\ref{Eq: Delta}), that fit the DAMA/LIBRA signal and do not supersaturate the null searches at $2\sigma$ for a benchmark point of $M_{\pid}=70$ GeV are shown in dark green (90\%C.L.) and light green (99\%C.L.).}
\end{figure}

Fig.~\ref{Fig: Banana} shows  a benchmark mass of $M_\pid=70 \GeV$ with  $90  \keV \le \Delta\le 150 \keV$ and  $50 \GeV \le f_\eff \le 400 \GeV$. The $90\%$ and $99\%$ ``confidence level''  contours are plotted, corresponding respectively to $\chi^2\le \chi^2_0+9.3$ and $\chi^2\le \chi^2_0+15$, where $\chi^2_0$ corresponds to the best fit point.

We compute the $\chi^2$ using the 12 half-keVee bins of DAMA's modulated signal and reported error bars between $2-8$ keVee, as well as the combined high energy bin from 8-12 keVee. This model fits DAMA's reported rate and nuclear recoil spectrum remarkably well. We also include in the $\chi^2$ contraints from null searches summarized in Table~\ref{NullData}, where for each null experiment we take as the standard error the $2\sigma$ Poisson fluctuation over the number of observed events. Moreover, we require the predictions for the null experiments not to supersatute the observations at the 95\%C.L.

\section{Cosmology}

In the strongly coupled model of this letter, $\Psi_H$ and $\Psi_L$ annihilate too efficiently for their thermal abundance 
to account for dark matter. 
Consequently, an asymmetry must be generated for ``dark meson number,'' $n_M\propto n_H -n_L$. 
Dark mesogenesis could in principle be tied to the Standard Model's baryogenesis. 

The $\pid$ meson is a simple iDM candidate.  
Here we show that a dominant fraction of the dark meson number asymmetry resides in $\pid$ mesons  
rather than exotic mesons, baryons, or $\rhod$ mesons. 

Exotic mesons and baryons are seeded by Coulomb-like bound states of heavy quarks $\Psi_H$ formed before confinement,
or created by merging of $\pid$ mesons after confinement. 
 
Before confinement, the gluon entropy exponentially suppresses 
$\Psi_H$ bound-state formation down to a temperature  $T^*\simeq E_B/\ln(s/n_{\text{M}})\sim E_B/30$, where $s$ is the entropy density of the Universe. For $N_c \ge 4$ and $M_\pid \lsim\OO(10^4)\Lambda$, $T^* \lsim \Lambda$ 
so gluon entropy prevents $\Psi_H$ bound-state formation with $N_H\geq 2$ until confinement. 

At confinement, $\Psi_L$ quark-antiquark pairs nucleate to screen the color charge of the $\Psi_H$. 
Confinement preferentially leads to the formation of 
$N_H=1$ dark meson over higher-$N_H$ dark mesons or baryons.
High-$N_H$ dark meson formation is negligible because the $\Psi_H$ are dilute at the time of confinement.
Formation of $N_H=1$ dark baryons is Boltzmann-suppressed for $N_c \gsim 4$
because they have $(N_c-2)$ more $\Psi_L$ constituents than dark mesons, so they are heavier. 

After confinement, heavy-quark binding occurs
via $\pid+\pid\rightarrow \pid^{(2)}+G$, where $G$ is a glueball and $\pid^{(2)}$ is an  $N_H=2$ dark meson.  
For $\Lambda\sim \OO(1\GeV)$  and $m_H \sim \OO(100\GeV)$,  these reactions are {\em endothermic} because the glueballs have a mass $m_G=\OO(\Lambda)\gsim E_B$.
The binding reactions of $\Psi_H$ require large momentum transfer $p_{\text{min}} \sim \sqrt{m_\pid (m_G -E_B)}\gg \Lambda$,
so the binding cross-section is controlled by perturbative $\Psi_H$ dynamics.
The thermally averaged $\pid$ binding cross-section is parametrically
\begin{eqnarray}
 \langle \sigma v\rangle \sim e^{-(m_G-E_B)/T}\frac{\alpha_t^2(p_{\text{min}})}{N_c^2m_H^2},
\end{eqnarray}
which is Boltzmann-suppressed for endothermic binding reactions. 

To summarize, $\Psi_H$ binding is suppressed by entropy for $T\leq\Lambda$,  
and by the endothermic Boltzmann factor for $T\geq\Lambda$.
In fact, $N_H\ne 1$ dark mesons have spin-0 ground states and similar scattering properties to $\pid$, 
so their abundances can be significant. 
Constraints arise only from $N_H=N_c$ dark baryons, with potentially large elastic scattering cross-sections.
Baryon formation proceeds through a sequence of $\Psi_H$-binding reactions,  
so a mild suppression of the binding rate at each stage significantly 
suppresses baryon formation.
This is discussed in detail in \cite{Alves:2010dd}.

The nearly degenerate $N_H=1$ meson spin states $\pid$ and $\rhod$ are equally populated at high temperatures.  
If the $\rhod$ decays only through kinetic mixing with hypercharge, the only kinematically allowed decays are to $\pid$ plus photons 
or neutrinos. These decays have lifetime longer than the age of the Universe \cite{Finkbeiner:2009}. 
Constraints from direct detection of $\rhod\rightarrow \pid$ de-excitation in nuclear scattering imply a 
fractional number-density bound $n_{\rhod}/n_{\pid} \lsim 10^{-2}$\cite{Finkbeiner:2009}.  
This constraint is endemic to all iDM models coupled to the Standard Model only through kinetic mixing, and is
troublesome if kinetic decoupling of dark matter occurs before $T\simeq 100 \keV.$
However, in CiDM models, $\rhod$ is de-excited through $\rhod+\rhod\rightarrow \pid+\pid$ scattering, 
with a large cross section set by the size of the dark meson, $\langle\sigma_{\rhod\rhod\rightarrow \pid\pid} v\rangle \simeq \Lambda^{-2}$.
For $M_\pid \sim 100 \GeV$ and $\Lambda \sim 1 \GeV$,
this reaction freezes out when $\exp(-\Delta/T_{\text{spin}}) \equiv n_{\rhod}/n_{\pid}\lsim 10^{-3}$, 
where $T_{\text{spin}}\lesssim \Delta$ is the asymptotic spin temperature.
After structure formation begins, up-scattering of $\pid$ into $\rhod$ can occur in dark matter halos, 
but the $\pid-\pid$ scattering is endothermic and the rate too small to re-populate the $\rhod$ state to
an observable level today.

Even though the CiDM self-scattering cross-section is large enough to depopulate the $\rhod$ abundance, it is nevertheless consistent with current bounds on dark-matter self-interactions (see Table I of \cite{Buckley:2009in})
\begin{eqnarray}
\nonumber
\frac{\sigma}{m}\simeq2\times10^{-6}\frac{\text{cm}^2}{\text{g}} \left(\frac{\GeV}{\Lambda}  \right)^2 \left(\frac{100\GeV}{M_\pid}  \right).
\end{eqnarray}
That is safely beneath the strongest present limits of $\sigma/m \lsim 10^{-2} \text{ cm}^2/\text{g}$. 



\section{Discussion}

The CiDM framework implements iDM in a parametrically novel manner.  
There are many generalizations of the model proposed in this letter -- for example, baryons and
weakly coupled bound states (``atoms'') \cite{Kaplan:2009de}. Many of these generalizations also naturally posses 
large cross sections to de-excite the $\OO(100 \keV)$ excited state in the early Universe\cite{Finkbeiner:2009}.
In contrast to weakly interacting iDM models, strongly interacting CiDM naturally leads to low spin temperatures 
and avoids de-excitation constraints. 

The recoil spectrum predicted by (\ref{eq:SpinDipoleScattering}) differs from conventional 
iDM spectra by a factor $\propto E_R$, offering a potential means of discriminating CiDM from 
iDM through direct detection. Moreover, specific CiDM models predict the existence of dark matter 
sub-components that arise from the small fraction of $\pid$ mesons that do process into other dark hadrons. 
Detection of these dark matter sub-components provides a striking signature of the underlying 
dynamics of CiDM \cite{Alves:2010dd}.  

The $U(1)_{A'}$ discussed in this letter has purely axial vector couplings to the dark quarks.    
Other charge assignments give rise to qualitatively different direct detection signatures, such as admixtures of elastic
and inelastic scatterings.  
Many of these alternate charge assignments generate elastic transitions of dark matter with distinctive nuclear
recoil spectra in addition to the inelastic transition \cite{Lisanti:2009vy}.

DAMA prefers $f_\eff\simeq 300$ GeV, resulting in $m_{A'} \lsim 30$ GeV, well beneath the energy frontier.
Searches at  high-intensity  $e^+e^-$ machines for CiDM through $A'$ production 
of light $\Psi_L$ matter may confirm or refute this entire class of models \cite{Essig:2009}.  
The coming generation of collider searches and direct detection experiments
will determine the true nature of DAMA's signal, possibly leading to irrefutable 
signatures of dark matter, and may unveil a whole new sector of physics.

\section*{Acknowledgements}
We thank Rouven Essig and Natalia Toro for numerous illuminating discussions, and Mariangela Lisanti for collaboration in setting constraints.
We also thank  Savas Dimopoulos, Michael Peskin, and Neal Weiner for helpful feedback.
SRB, PCS  and JGW are supported by the US DOE under contract number DE-AC02-76SF00515.
DSMA is supported by the NSF under grant PHY-0244728.


\end{document}